\documentclass[%
 superscriptaddress,
 reprint,
 showpacs,preprintnumbers,
 nofootinbib,
 amsmath,amssymb,
 aps,
 prl, %prd
 floatfix,
]{revtex4-1}

\usepackage[table,xcdraw]{xcolor}

\usepackage{amsmath}
\usepackage{amssymb}
\usepackage{float}

\usepackage{graphicx}% Include figure files
\usepackage{dcolumn}% Align table columns on decimal point
\usepackage{bm}% bold math
\usepackage{afterpage}
\usepackage{tikz}
\usepackage{textgreek}
\usepackage{caption}
\RequirePackage[colorlinks,citecolor=blue,urlcolor=blue,linkcolor=blue]{hyperref}
\usepackage{todonotes}

\usepackage{xspace}

\newcommand{\affiliationCERN}{European Organization for Nuclear Research (CERN), 1211 Geneva 23, Switzerland}
\newcommand{\affiliationUCh}{University of Chicago, Chicago, IL 60637, USA}

\def\bracketbar{\smash{\hbox{\kern-7pt\raise3pt%
    \hbox{{\tiny(}{\lower1.5pt\hbox{\bf--}}{\tiny)}}}}}
\providecommand{\varpm}{\mathbin{\vcenter{\hbox{%
  \oalign{\hfil$\scriptstyle+$\hfil\cr
          \noalign{\kern-.3ex}
          $\scriptscriptstyle({-})$\cr}%
}}}}
\providecommand{\varmp}{\mathbin{\vcenter{\hbox{%
  \oalign{$\scriptstyle({+})$\cr
          \noalign{\kern-.3ex}
          \hfil$\scriptscriptstyle-$\hfil\cr}%
}}}}

\begin{document}

\title{The role of final-state interaction modeling in neutrino energy reconstruction and oscillation measurements}% Force line breaks with \\

	\author{Y.~Liu}
	\email[Contact e-mail: ]{yinrui@uchicago.edu}
	\affiliation{\affiliationUCh}

	\author{L.~Munteanu}
	\email[Contact e-mail: ]{laura.munteanu@cern.ch}
	\affiliation{\affiliationCERN}

	\author{S.~Dolan}
	\email[Contact e-mail: ]{stephen.joseph.dolan@cern.ch}
	\affiliation{\affiliationCERN}

\date{\today}% It is always \today, today,

\begin{abstract}
We present a quantitative demonstration that, without additional theoretical and experimental efforts, realistic variations in final-state-interaction (FSI) modeling may alter reconstructed neutrino-energy spectra at next-generation long-baseline experiments by amounts comparable to, or larger than, variations induced by oscillation-parameter shifts at their projected precision. Using the DUNE flux and baseline as a case study, we show that these FSI-driven distortions can mimic the effects of changes in the oscillation parameters $\Delta m^2_{32}$ or $\delta_{\rm CP}$, producing a potential degeneracy. Our analysis thereby underscores the urgent need for an improved characterisation of FSI to enable robust constraints from near detectors through the development of theory-driven uncertainty parameterisations benchmarked with dedicated new measurements.

\end{abstract}

\maketitle

%%%%%%%%%%%%%%%%%%%%%%%%%%%%%%%%%%%%%%%%%%%%%%%%%%
\textit{Introduction} ---
%\begin{itemize}
%    \item Very short intro to LBL
%    \item Importance of enu rec, especially for DUNE
%    \item Relevance of FSI and potential degeneracies with osc parameters 
%    \item In this manuscript we ... 
%\end{itemize}
Long-baseline (LBL) neutrino oscillation experiments are essential for advancing our understanding of neutrino properties through measurements of $\nu^{\bracketbar}_{\mu}$ disappearance and  $\nu^{\bracketbar}_e$ neutrino appearance in an  $\nu^{\bracketbar}_{\mu}$ neutrino beam. Future experiments, the Deep Underground Neutrino Experiment (DUNE)~\cite{DUNE:2020ypp} and Hyper-Kamiokande (HK)~\cite{Hyper-Kamiokande:2018ofw,Hyper-Kamiokande:2025fci}, will dramatically improve measurements on neutrino oscillation parameters, determine whether there is charge-parity (CP) violation in the lepton sector, resolve the neutrino mass ordering, and search for signatures of new physics~\cite{duneCDRphys, Hyper-Kamiokande:2025fci}.

A vital component of these measurements is the precise reconstruction of the incoming neutrino energy from the visible products of neutrino interactions with the nuclear targets within detectors~\cite{NuSTEC:2017hzk}. This is particularly important for experiments like DUNE that are designed with a broadband neutrino beam and will infer oscillation probabilities over a wide energy range. The core technology of the DUNE detectors is the liquid argon time projection chamber (LArTPC), which can provide a detailed reconstruction of neutrino interaction products following neutrino interactions with argon nuclei~\cite{DUNE:2020ypp}. The reconstruction of the incident neutrino energy relies on summing the energy of the visible interaction products. Since different particles leave different fractions of their energy as visible signatures in a LArTPC, the accuracy of neutrino energy reconstruction depends on the modeling of the constituents of the hadronic final state. This is sensitive to final-state interactions (FSI), where hadrons produced in the initial neutrino interaction undergo further scattering inside the nucleus before exiting, altering observable particle multiplicities and kinematics~\cite{Dytman:2021ohr,NuSTEC:2017hzk} . 

%FSI can alter the kinematics and even change the type of particles that can be detected in the final state of neutrino interactions, leading to biases and uncertainties in the inferred neutrino energy~\cite{NuSTEC:2017hzk}. These biases pose concerns as they can obscure oscillation effects, potentially causing degeneracies with oscillation parameters. Furthermore, there are large uncertainties associated with the models that describe FSI processes~\cite{Dytman:2021ohr}.

In this work, we use simulations of neutrino interactions with different FSI models with the DUNE neutrino flux~\cite{dunefluxurl,DUNE:2020ypp} and baseline (1300 km) as a case study to assess how different FSI models affect reconstructed neutrino energies and the resulting energy spectra. We then provide a novel comparison of the size of these distortions to variations of oscillation parameters at the level of precision that future LBL experiments aim to measure~\cite{Hyper-Kamiokande:2025fci, DUNE:2020jqi}. %Through this study, we highlight the potential for realistic variations of FSI modeling to be degenerate with variations of neutrino oscillation parameters, thereby emphasizing the need for precise modeling of FSI to achieve the experiments' ambitious goals.

%%%%%%%%%%%%%%%%%%%%%%%%%%%%%%%%%%%%%%%%%%%%%%%%%%
\textit{Method and models} ---
%\label{sec:methods}
%The goal of the analysis is to evaluate how the expected reconstructed spectrum of neutrino energies at the DUNE far detector baseline varies with alterations to FSI modeling and how this compares to the variations of neutrino oscillation parameters. 
Charged-current neutrino interactions with argon are generated using the \texttt{GENIE v3.04.00} neutrino event generator~\cite{Andreopoulos:2015wxa, Andreopoulos:2009rq} using a fixed neutrino-nucleus interaction model (described below) apart from alterations to the simulation of FSI. Electron and muon neutrino and antineutrino interactions are generated using fluxes expected at the DUNE near detector~\cite{dunefluxurl,DUNE:2020ypp} and then weighted to apply oscillations using the \texttt{Prob3++} software package~\cite{Barger:1980tf,prob3pp}. The baseline set of neutrino oscillation parameters considered are $\sin^2\theta_{12}=0.307$, $\sin^2\theta_{13}=0.0219$, $\sin^2\theta_{23}=0.558$, $\Delta m^2_{21}=7.53\times10^{-5}$ eV$^2$, $\Delta m^2_{32}=2.455\times10^{-3}$ eV$^2$, taken from Ref.~\cite{ParticleDataGroup:2024cfk}, $\delta_{\rm CP}=0$ by default, and a normal neutrino mass ordering. Matter effects are treated using the default implementation within \texttt{Prob3++}~\cite{Barger:1980tf,prob3pp}.

The baseline interaction model uses the \texttt{G18\_10a\_00\_000} \texttt{GENIE} configuration~\cite{GENIE:2021zuu}, employing a local Fermi-gas nuclear ground state, the Valencia group's model for CCQE and 2p2h interactions~\cite{Nieves:2011pp, gran2013neutrino, Schwehr:2016pvn}, the Berger-Seghal pion production models~\cite{Berger:2007rq,PhysRevD.77.059901}, and GRV98 PDFs with Bodek-Yang corrections ~\cite{Gluck:1998xa,Bodek:2002vp} for DIS. Hadronization is modeled either with  PYTHIA~\cite{SJOSTRAND199474,Sjostrand:2006za} (at invariant masses, $W$, above 3.0 GeV/$c^2$) or the custom AGKY model~\cite{Yang:2009zx} ($W<2.3$ GeV/$c^2$), and an interpolation between them~\cite{GENIE:2021wox}. 

FSI are simulated using one of the four different models available within GENIE: the default hA2018 model, the hN2018 model~\cite{GENIE:2021zuu}, the Liege Intranuclear Cascade model (INCL)~\cite{Boudard:2012wc}, or the GEANT4 Bertini Cascade model (G4BC)~\cite{Heikkinen:2003sc}. Very broadly, hA2018 and hN2018 are widely-used models that are tuned extensively to hadron scattering data, while INCL and G4BC are more sophisticated hadron transport models. Further details and comparisons of these FSI models are available in the Supplemental Material and in Refs.~\cite{Dytman:2021ohr, ershova2022study}. The span of these models provides a realistic \textit{minimal} uncertainty envelope.

In this work, we define a proxy for reconstructed neutrino energy ($E^{\mathrm{rec}}_{\nu}$) as:
\begin{equation}
    E^{\mathrm{rec}}_{\nu} =  E_{\ell} + \sum_{i=p}{T_i} + \sum_{i=\pi, K, \gamma}{E_i}
    \label{eq:enuhad}.
\end{equation}
This is the sum of the lepton energy ($E_\ell$), the kinetic energies ($T$) of all final-state protons, and the total energy ($E$) of all pions, kaons (charged or neutral) and photons. No selection effects or energy thresholds are considered. We define the relative neutrino energy bias as $(E^{\mathrm{rec}}_{\nu}-E^{\mathrm{true}}_{\nu})/E^{\mathrm{true}}_{\nu}$.  

\autoref{eq:enuhad} assumes a neutrino detector which acts as a perfect calorimeter for charged-particles and photons whilst also successfully identifying the multiplicity of charged pions\footnote{Charged pion mass energy is lost largely to neutrinos from the pion decay, and thus it must be added to the calorimetric sum.}. In this definition, the size of the bias is primarily driven by the amount of the interaction energy carried away by neutrons\footnote{Other small contributions to bias may stem from nuclear binding energy.}.  This bias metric is broadly illustrative of what might be expected from DUNE's LArTPCs, which can use a combination of tracking and calorimetry to achieve this high precision for neutrino energy reconstruction. The relationship between this proxy and the true neutrino energy depends on the modeling of effects which modify the visible energy in the detector, like FSI. Ref.~\cite{Wilkinson:2022dyx} provides a discussion of this reconstructed neutrino energy metric and how it compares with others.  %\autoref{eq:enuhad} is not applicable to the case of water Cherenkov detectors, which are used for HK, as these are not sensitive to most hadrons below Cherenkov threshold.
The impact of detector smearing and reconstruction thresholds are studied in the Supplemental Material and found not to change the conclusions of this work.

\begin{figure}[!htb]
\centering
\includegraphics[width=0.40\textwidth]{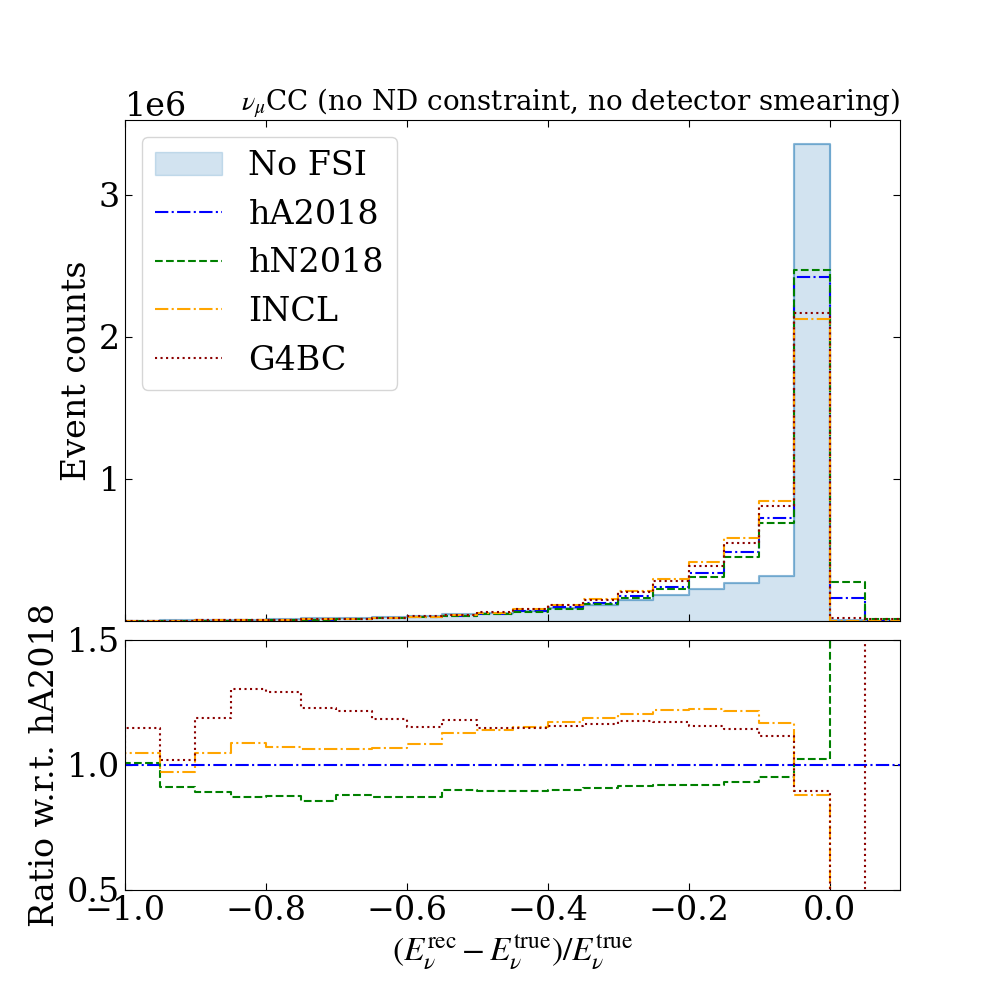}
\includegraphics[width=0.40\textwidth]{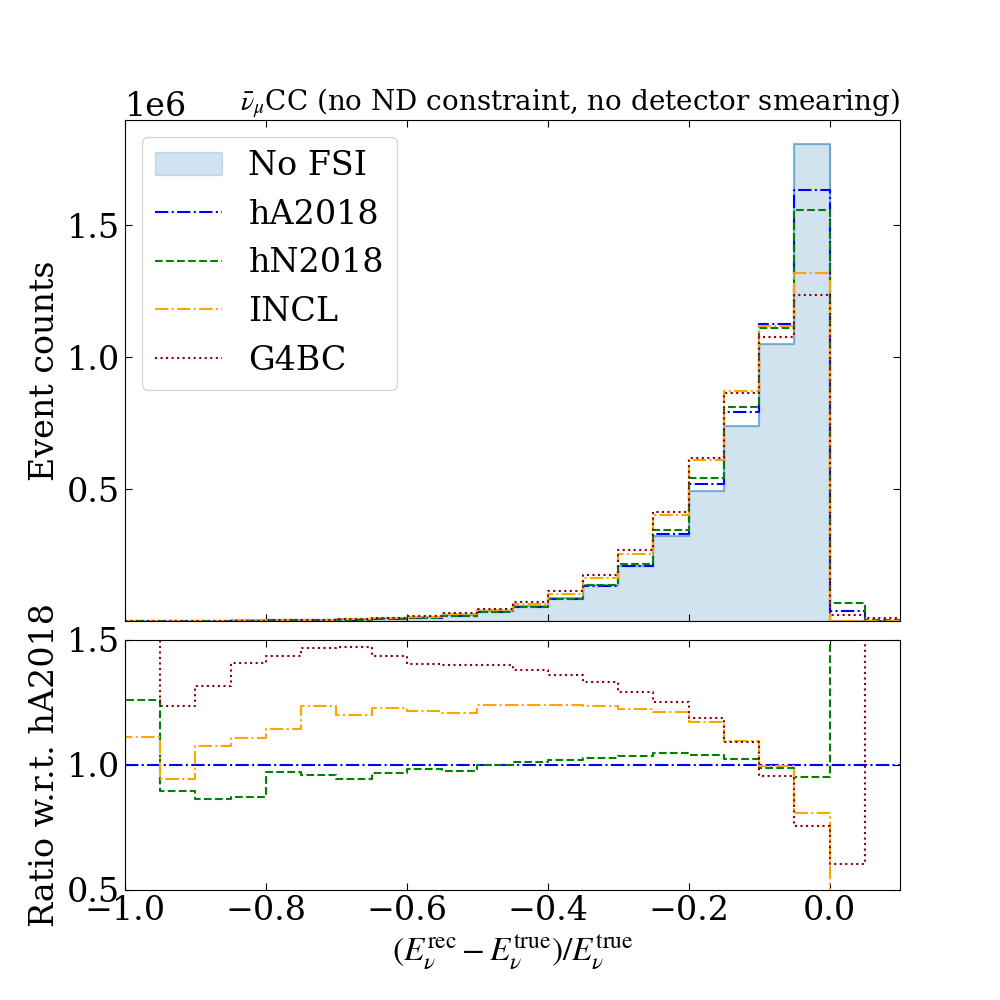}
\caption{The simulated distribution of relative neutrino energy bias for 5 million muon neutrino (top) and antineutrino (bottom) charged-current interactions, using the DUNE neutrino and antineutrino beam-mode fluxes respectively before oscillations occur, for a variety of different FSI models. The blue shaded histogram assumes no FSI, while the other histograms correspond to the four FSI models considered in this work. The bottom panel shows the ratios of different FSI model predictions with respect to the hA2018 model. \vspace{-4mm}} 
\label{fig:nurecdemo}
\end{figure}

\autoref{fig:nurecdemo} shows the simulated distribution of the relative neutrino energy bias for $\nu^{\bracketbar}_{\mu}$ charged-current interactions for each of the different FSI models, including one simulation without FSI. It is apparent that FSI has a significant impact on neutrino energy reconstruction and that different FSI models lead to significantly different energy bias profiles. Moreover, we note that FSI, and the sources of invisible energy it introduces, has a different impact on the energy reconstruction of neutrinos with respect to antineutrinos. Whilst neutrino interactions occur primarily with neutrons in the nucleus, and as a result often produce protons or other charged particles in the absence of FSI, antineutrinos reverse this behavior and produce more neutrons in the final state. This can be observed in the much broader distribution of the relative bias in the antineutrino case. %As a reference, for the models considered, the fraction of events with a bias larger than 10\% is of order 40\% for neutrinos and 50\% for antineutrinos.

Variations to the reconstructed neutrino energy spectra for $\nu^{\bracketbar}_{\mu}$ disappearance and $\nu^{\bracketbar}_{e}$ appearance channels from alterations of FSI modeling are compared to variations of neutrino oscillation parameters. The latter are chosen to be comparable to next-generation experiments' long-term projected precision~\cite{DUNE:2020jqi,Hyper-Kamiokande:2025fci}. For reference, these are also compared to expected statistical uncertainties on the measured reconstructed neutrino energy spectra from a projected exposure of ten years of the DUNE experiment (corresponding to 624 kt-MW-years, equally split between beam operation in neutrino and antineutrino running modes~\cite{DUNE:2020jqi}\footnote{The total signal statistics have been obtained by scaling the signal statistics reported in~\cite{DUNE:2020jqi} by the ratio between 336 kt-MW-years (7 years) and 624 kt-MW-years (10 years) exposures.}). In these studies, we consider that there is no background contribution from wrong-sign or wrong-flavor neutrino candidates or neutral-current interactions.

\begin{figure}[!htb]
\centering
\includegraphics[width=0.40\textwidth]{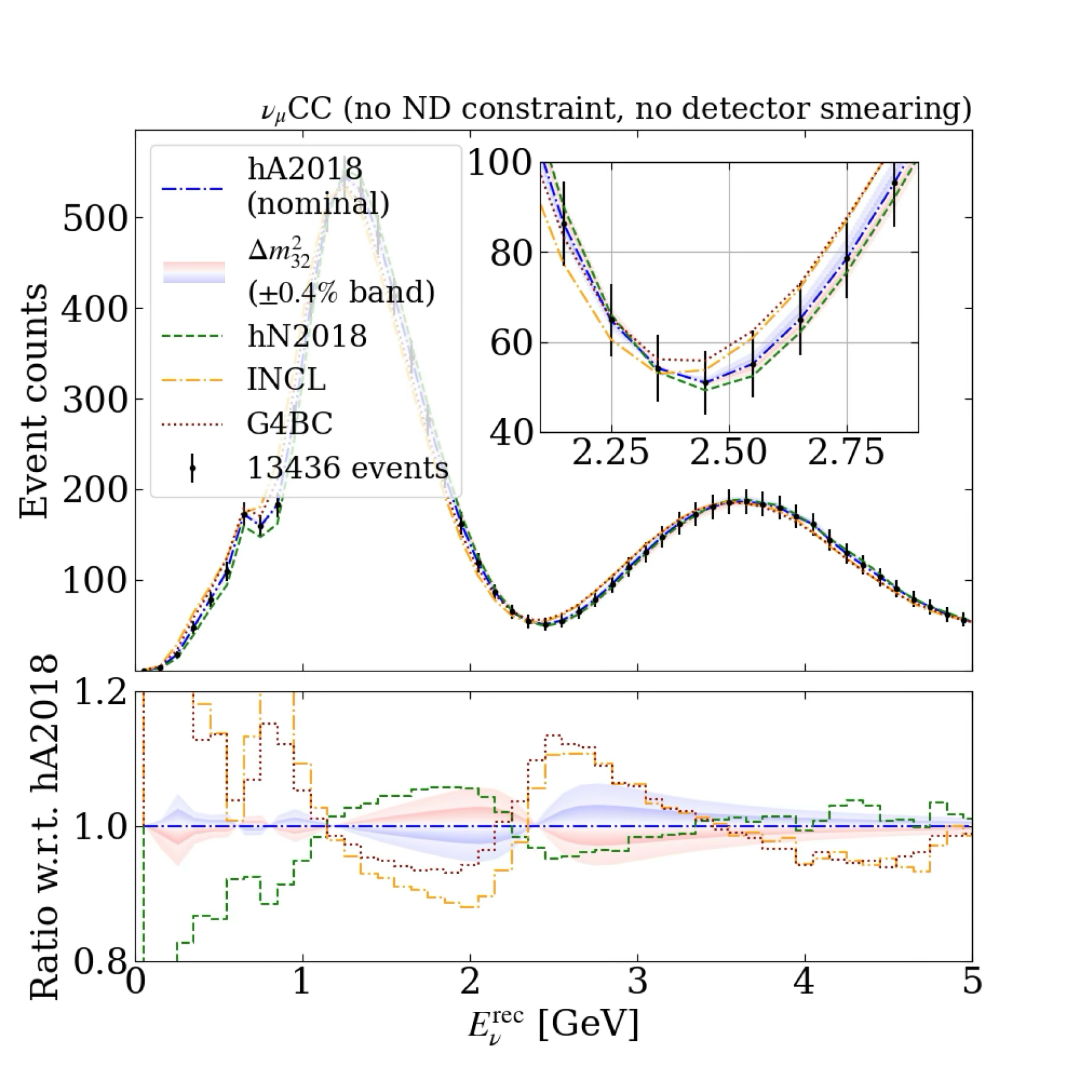}
\includegraphics[width=0.40\textwidth]{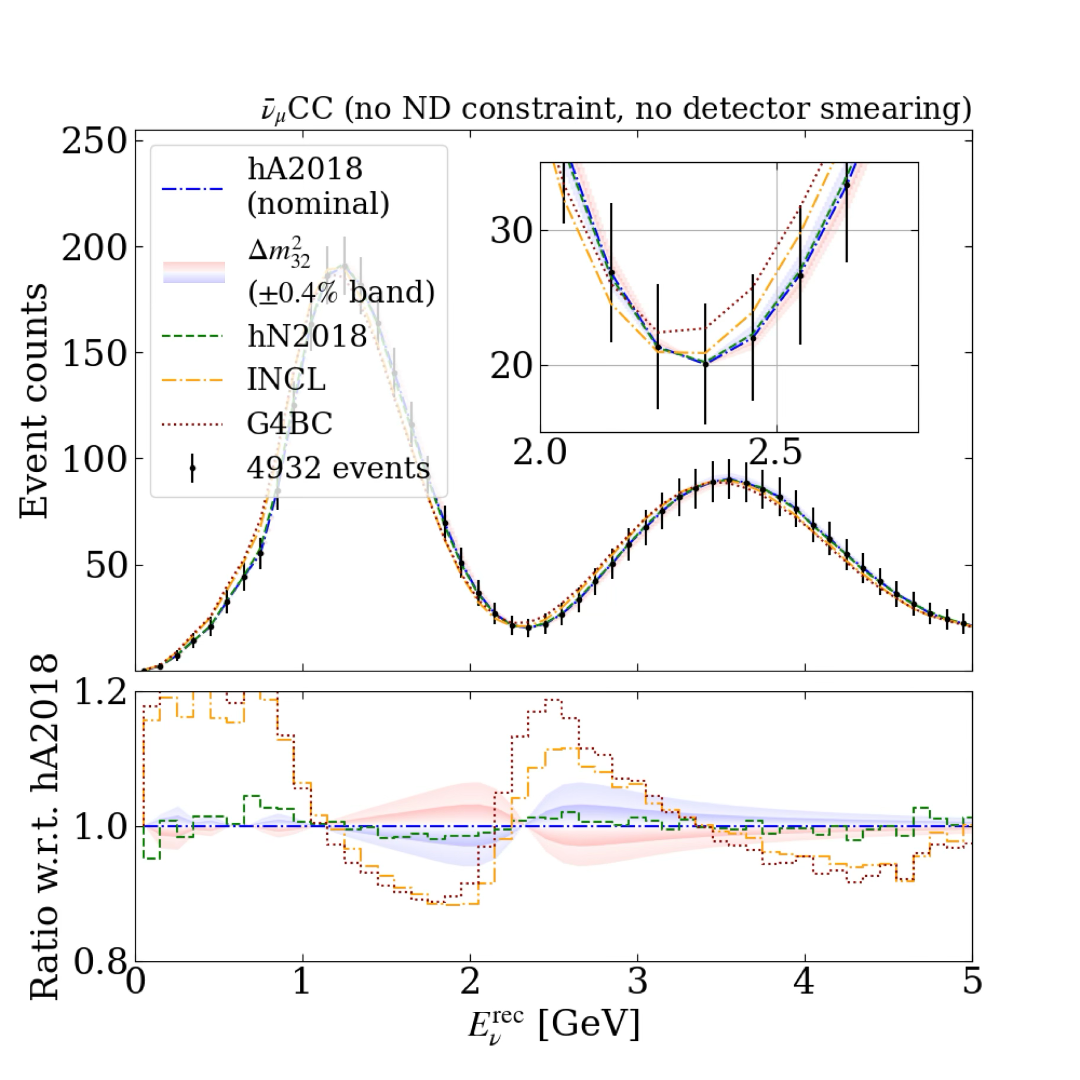}
\caption{Comparison of the simulated $\nu^{\bracketbar}_{\mu}$ reconstructed energy spectra and an oscillated DUNE experiment flux using different FSI models, to variations of $\Delta m_{32}^2$. This is simulated for neutrino (top) and antineutrino (bottom) beam modes. In each sub-figure, reconstructed energy spectra under different FSI models are shown as histograms and $\pm 0.4$\% $\Delta m_{32}^2$ variations are shown with shaded red (+) and blue (-) bands centered on the hA2018 baseline. The error bars are broadly indicative of the DUNE experiment's statistical uncertainty with 10 years of operation. The inset provides a zoomed-in view of the first oscillation maximum. The lower panel displays ratios of FSI model and $\Delta m_{32}^2$ variations relative to the hA2018 baseline. In the lower panel, the bands show 1~$\sigma$ (inner) and 2~$\sigma$ (outer) variations to $\Delta m_{32}^2$. \vspace{-4mm}}
%Impact on disappearance dm32 measurements. The sensitivity used current PDG value and error. The statistics used \href{https://arxiv.org/pdf/2006.16043}{DUNE paper} Table 7 scaled by 624/336 from 7-year to 10-year exposure. Error bands indicate 1- and 2-sigma.} 
\label{fig:numu_dm32}
\end{figure}

%%%%%%%%%%%%%%%%%%%%%%%%%%%%%%%%%%%%%%%%%%%%%%%%%%
\textit{Results} ---
%\begin{itemize}
%    \item Compare variations for numu disappearance / dm32 in \autoref{fig:numu_dm32}
%    \item Compare variations for nue disappearance / dcp in \autoref{fig:nue_dcp}
%    \item Add additional figure showing variations to dcp DUNE hopes to measure rather than just looking at CPC vs CPV
%    \item Other figures we don't show (e.g. theta 23) are in an appendix / supplementary material
%\end{itemize}
We first examine the effect of FSI modeling on reconstructed energy spectra relevant to $\nu^{\bracketbar}_{\mu}$ disappearance analyses, in both neutrino and antineutrino beam modes. \autoref{fig:numu_dm32} shows the reconstructed oscillated spectra using different FSI models, overlaid with $\pm 0.4$\% variations of $\Delta m_{32}^2$.
%within its current uncertainty as reported by the PDG~\cite{ParticleDataGroup:2024cfk}.
As indicated in the bottom panel in each sub-figure, the shifts induced by FSI variations are comparable in size to the simulated variations of $\Delta m_{32}^2$, and reproduce characteristic features of oscillation parameter variations near the first oscillation maximum corresponding to a shift in the energy spectrum. Additional figures comparing the FSI effects with corresponding variations in $\sin^2\theta_{23}$ are provided in the Supplemental Material. 

\begin{figure*}[htpb]
\centering
\includegraphics[width=0.40\textwidth]{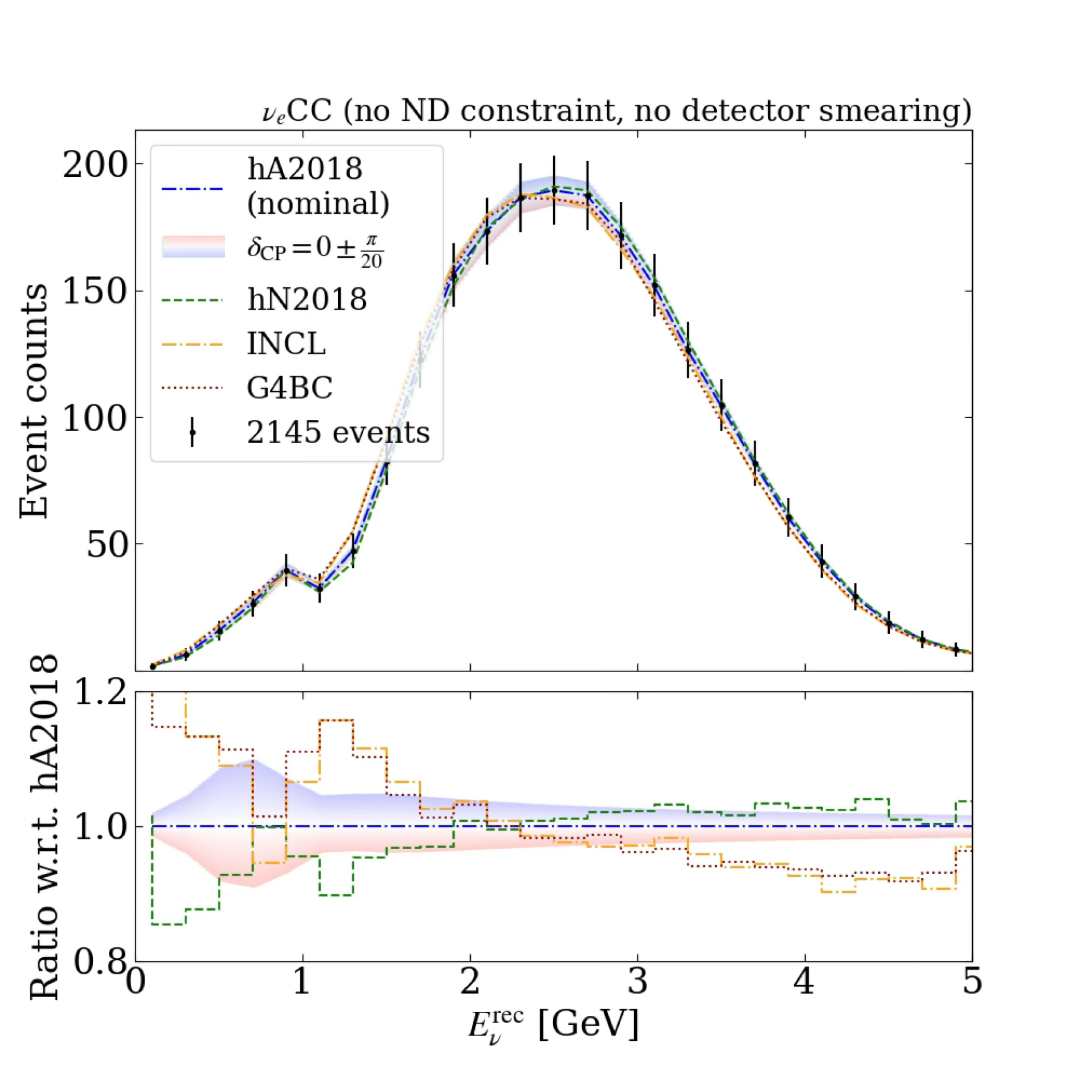}
\includegraphics[width=0.40\textwidth]{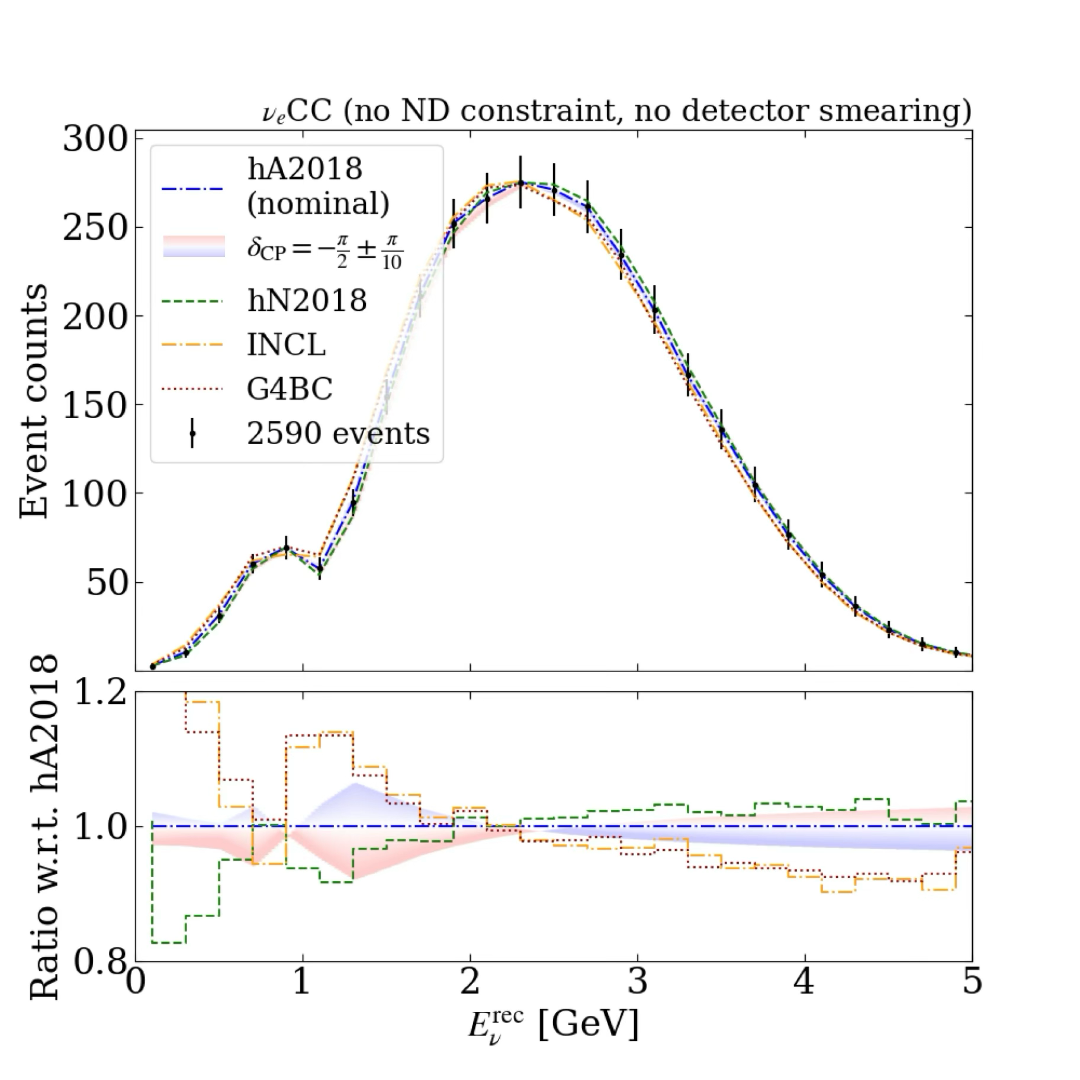}

\includegraphics[width=0.40\textwidth]{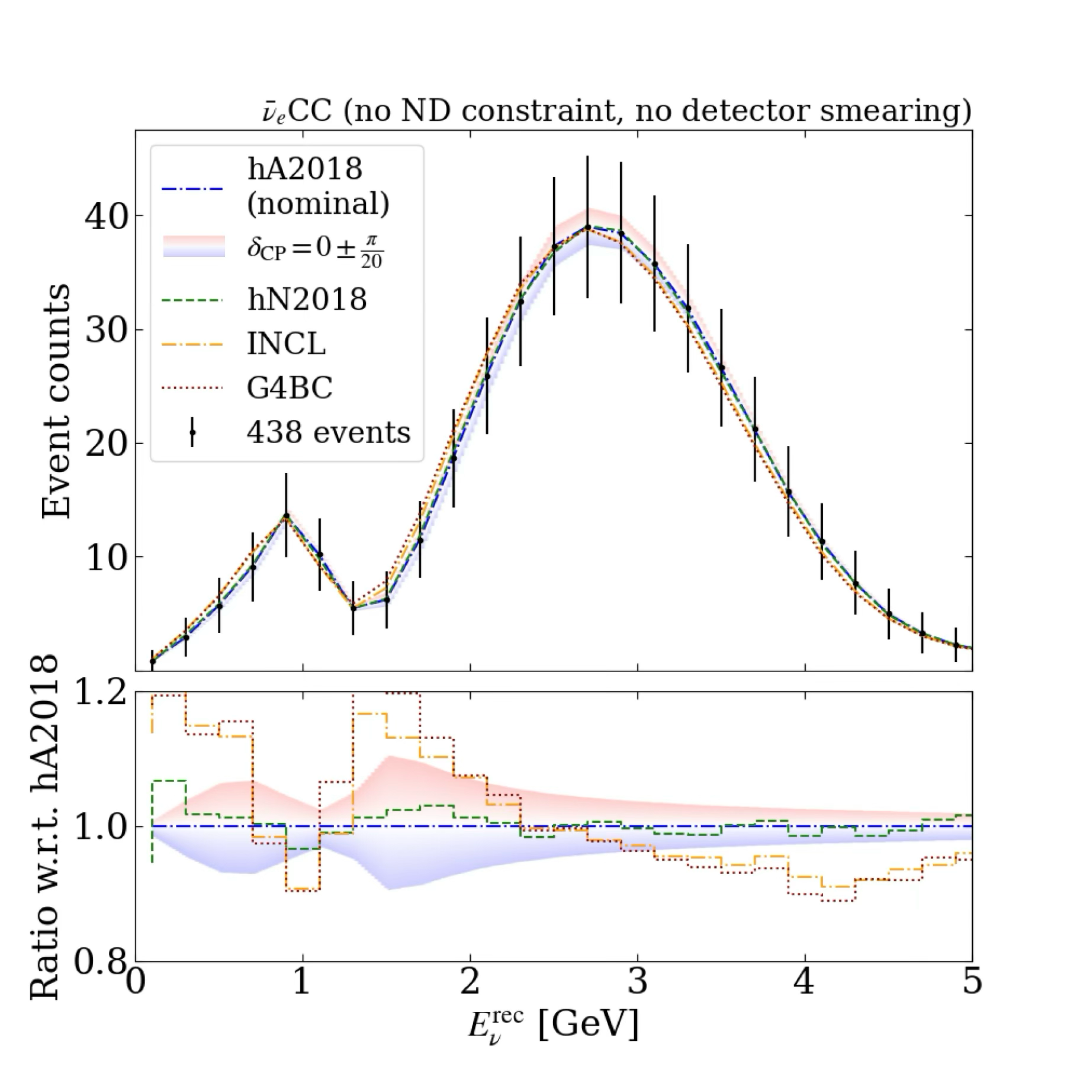}
\includegraphics[width=0.40\textwidth]{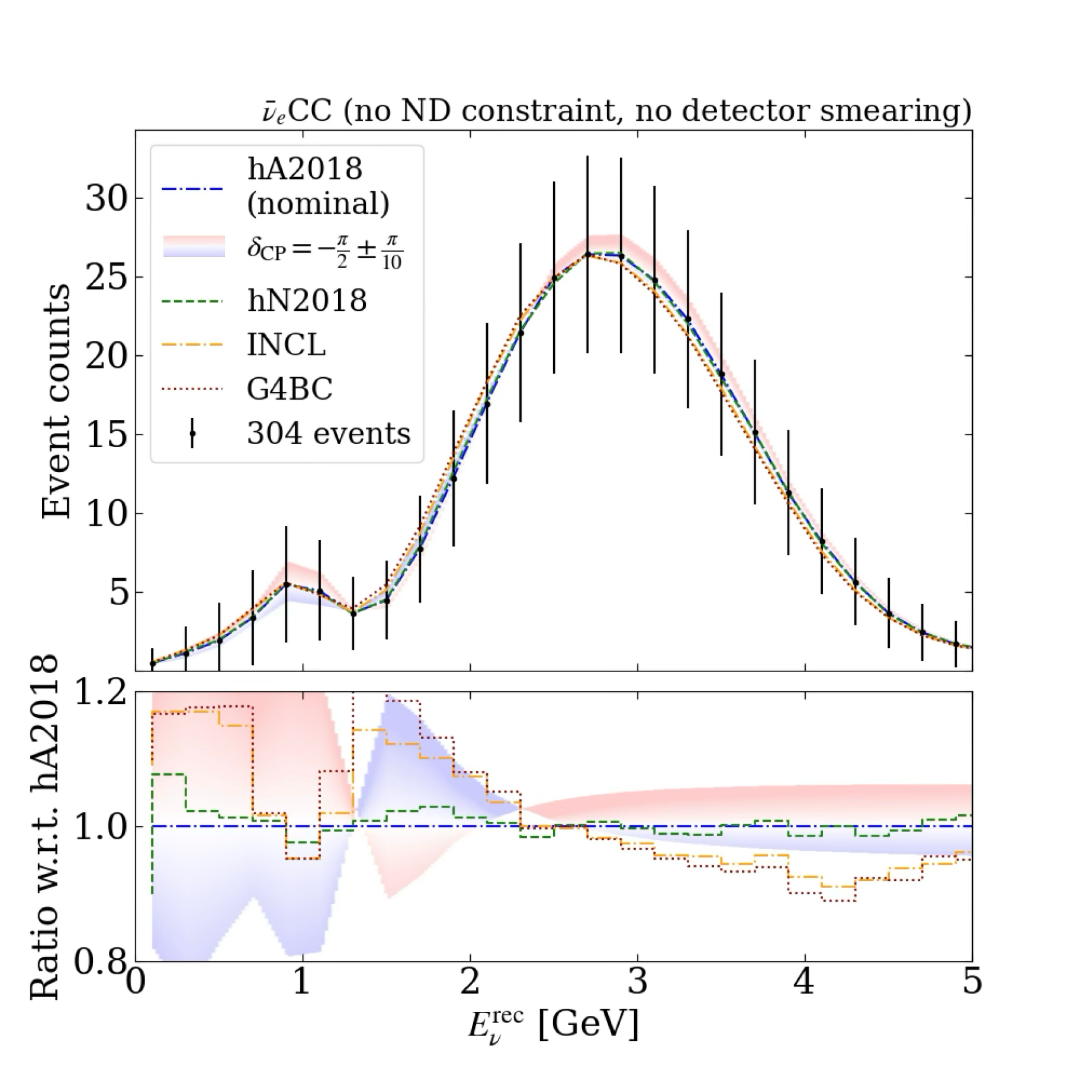}
\caption{Comparison of the simulated $\nu^{\bracketbar}_{e}$ reconstructed energy spectra and an oscillated DUNE experiment flux using different FSI models, to variations of $\delta_{\rm CP}$. This is simulated for neutrino (top) and antineutrino (bottom) beam modes. The left and right correspond to $\delta_{\rm CP}=0$ and $\delta_{\rm CP}=-\pi/2$, respectively. Other figure elements follow the same conventions as described in \autoref{fig:numu_dm32}, except that only the 1~$\sigma$ uncertainty band is shown in the lower panels. \vspace{-4mm}}

\label{fig:nue_dcp}
\end{figure*}

A similar analysis is performed for the $\nu^{\bracketbar}_{e}$ appearance channel, which drives the sensitivity for identifying CP violation (CPV) and the precise measurement of the CP-violating phase $\delta_{\rm CP}$. \autoref{fig:nue_dcp} shows the reconstructed $\nu_e$ and $\bar{\nu}_e$ spectra under the CP conservation ($\delta_{\rm CP}=0$) and maximal CPV ($|\delta_{\rm CP}|=\pi/2$) hypotheses. When referring to maximal CPV across the study, we choose to only consider the case where $\delta_{\rm CP}$ is -$\pi/2$, which is closest to the current knowledge~\cite{ParticleDataGroup:2024cfk}. Since sensitivity to measuring $\delta_{\rm CP}$ depends on its value, this is shown alongside an error band of $\pm \pi/10$ or $\pm \pi/20$ for the CP-conserving and CP-violating cases respectively, approximately corresponding to the projected precision reported in Refs.~\cite{DUNE:2020jqi, Hyper-Kamiokande:2025fci}. Across all four panels, the variations from FSI modeling are comparable to, or exceed, the size of spectral differences induced by varying $\delta_{\rm CP}$. However, as can also be seen by comparing the integrated event rate of CP-conserving case to the CP-violating case, the observation of CPV in the $\nu^{\bracketbar}_{e}$ appearance channel is clearly robust against considered FSI modeling variations if CPV is maximal. Whilst not shown here, we also studied the sensitivities to $\theta_{13}$, the $\theta_{23}$ octant determination and the neutrino mass ordering, concluding that the considered FSI modeling uncertainty has minimal impact on these measurements.

%%%%%%%%%%%%%%%%%%%%%%%%%%%%%%%%%%%%%%%%%%%%%%%%%%
\textit{Discussion and outlook} ---
%%% FSI brings bias to neutrino energy reconstruction
\autoref{fig:nurecdemo} shows that FSI changes the magnitude and shape of the bias in our proxy for the neutrino energy reconstruction at DUNE. Changing the FSI model varies the number of (anti)neutrino events with more than 10\% relative bias by about 30\% (20\%). Whilst not explicitly shown here, the evolution of the bias with neutrino energy is also changed (see Supplemental Material).
The FSI-induced modification in this bias manifests as spectral shifts and smearing in $E^{\mathrm{rec}}_{\nu}$. This affects the \textit{measured} position of the first oscillation maximum, whose position as a function of \textit{true} neutrino energy is determined primarily by the value of $\Delta m^2_{32}$. This highlights a risk of degeneracy between FSI effects and $\Delta m^2_{32}$ variations. As shown in \autoref{fig:numu_dm32}, the model spread from the baseline can exceed the spectral shifts induced by a 0.4\% change in $\Delta m^2_{32}$, particularly for the G4BC and INCL simulations.

%%% FSI impacts $\Delta m^2_{32}$ sensitivity
%The FSI-induced modification in reconstructed neutrino energy bias manifests as a shift and smear of the measured neutrino energy spectra. This affects the \textit{measured} position of the first oscillation maximum corresponding to the minimum of $\nu^{\bracketbar}_{\mu}$ survival probability, whose position as a function of \textit{true} neutrino energy is determined primarily by the value of $\Delta m^2_{32}$. In \autoref{fig:numu_dm32}, the variation of reconstructed neutrino energy from FSI model spread around the measured first oscillation maximum is larger than that of a shift in $\Delta m^2_{32}$ by 0.4\%, the target precision of future experiments. This highlights the risk of a degeneracy between FSI effects and variations of $\Delta m^2_{32}$ in the data of far detectors for LBL experiments. For both the neutrino and antineutrinos cases, the most significant differences with respect to the baseline hA2018 model are exhibited by the sophisticated G4BC and INCL models.

%%% FSI impacts $\delta_{\rm CP}$ sensitivity
In the case of true maximal CPV, our analysis indicates that the impact of FSI modeling is not likely to be decisive in the search for CPV since the majority of this sensitivity comes directly from the relative amount of observed electron neutrinos and antineutrinos rather than shape alterations in the reconstructed neutrino energy spectra. However, the situation is different for precision measurements of $\delta_{\rm CP}$. In this case, the measurement can be influenced by effects that alter the shape of the reconstructed electron (anti)neutrino energy spectrum, and the degree of sensitivity depends on the true value of $\delta_{\rm CP}$. As shown in \autoref{fig:nue_dcp}, when $\delta_{\rm CP}\sim-\pi/2$, variations in $\delta_{\rm CP}$ primarily manifest as shape changes in the reconstructed energy spectra, whereas for $\delta_{\rm CP}\sim0$, they appear more as overall rate differences.
Since the shape dependence resembles the effects introduced by variations in FSI modeling, we conclude that for precision $\delta_{\rm CP}$ measurements, FSI uncertainties may be significant if the true value of $\delta_{\rm CP}$ corresponds to near-maximal CPV.

%%% Limitations of the study
The results presented in this work offer a quantitative assessment of the impact of FSI modeling on $E^{\mathrm{rec}}_{\nu}$ but are not representative of a systematic uncertainty for a real oscillation analyses due to a number of limitations. First, we  consider variations between FSI models as a proxy for their potential effect, neglecting the effect they will have on any constraint propagated from a near detector. In the case of DUNE, the near detector is clearly well equipped to constrain FSI modeling uncertainties~\cite{DUNE:2021tad}, but doing so requires a currently unavailable comprehensive parameterisation of FSI to be fitted, alongside parameters describing other uncertainties, to the near detector data. The authors believe that the impact of this constraint cannot be meaningfully approximated with reasoning given in the Supplemental Material. Second, we only examine FSI variations on the size of the expected shape of the oscillated signal, neglecting its potential impact on backgrounds originating from sources such as wrong sign, wrong flavor or neutral current interactions. %We note that the inclusion of these background sources will also require a corresponding understanding of FSI processes. 
Third, the effect of FSI variations on the reconstructed neutrino energy spectrum is studied in isolation of other observables. It may be possible to choose an alternative set of far detector observables which could help disambiguate FSI effects from genuine neutrino oscillations (e.g. by using transverse or generalized kinematic imbalance variables~\cite{Lu:2015tcr,Lu:2015hea,Furmanski:2016wqo,MicroBooNE:2023krv}).\footnote{Regardless of the choice of observables, their relationship to the true neutrino energy remains a model-depending mapping which in turn must be accompanied by an appropriate uncertainty parameterisation and therefore an improved understanding of FSI.} Fourth, we do not account for backgrounds to far detector event selections from neutral current interactions, neutrinos of the wrong flavor or sign, and the intrinsic electron neutrino contamination. Whilst these backgrounds are small for neutrino analyses ($\lesssim$20\%~\cite{DUNE:2020jqi}) they can be large for antineutrino samples (up to $\sim$65\%~\cite{DUNE:2020jqi}). The size and shape of the backgrounds will themselves also be impacted by FSI. 

This work also does not consider potential variation of the reconstructed neutrino energy spectra from sources of uncertainty beyond FSI. Modeling of other nuclear effects, particularly those that can significantly alter the hadronic system (e.g. hadronization), and detector response gives scope for significantly more variation. Although we consider only two sets of oscillation parameters, we expect that the qualitative conclusions of the study will not change for choices within the ranges allowed by experimental measurements~\cite{ParticleDataGroup:2024cfk}.

%%% The role of ND
Despite the above limitations, our analysis demonstrates the potential impact that varying FSI alone can have on reconstructed neutrino energy spectra and compares it to oscillation parameter variations, showing scope for significant degeneracy between the two. Mitigating this will require future near detector systems to be able to precisely characterise FSI. In particular, the More Capable Near Detector (MCND)~\cite{DUNE:2024wvj} for DUNE Phase II will likely play a critical role in constraining interaction models through its enhanced ability to resolve final-state particles via low particle tracking thresholds. Moreover, the movable DUNE near detector system may be able to leverage the PRISM technique to remove oscillation measurement bias in FSI modeling in exchange for larger uncertainties~\cite{DUNE:2021tad}.

Beyond near detector constraints, resolving FSI-driven degeneracies requires concerted theoretical guidance and dedicated experimental efforts. While hadron scattering on argon remains poorly constrained~\cite{Dytman:2021ohr}, modern LArTPC experiments including LArIAT~\cite{LArIAT:2019kzd} and ProtoDUNE~\cite{DUNE:2017pqt,DUNE:2020cqd} have begun using test beams to measure hadron-argon cross sections~\cite{LArIAT:2021yix,DUNE:2024qgl,DUNE:2025pda,DUNE:2025zhx}. Although these measurements currently suffer from large uncertainties, future exclusive or differential analyses, potentially similar to carbon-target analyses from HADES~\cite{Hojeij:2023rsq, Ramstein:2024uea}, may offer significant discriminating power between different FSI models. These measurements should quantify the fraction of initial-state hadron energy visible in the final state, thereby constraining the shape of \autoref{fig:nurecdemo}. Furthermore, FSI is being probed across various energies and channels by neutrino experiments including MicroBooNE~\cite{MicroBooNE:2016pwy}, SBND~\cite{Machado:2019oxb}, ICARUS~\cite{ICARUS:2004wqc}, and the DUNE 2x2 prototype~\cite{DUNE:2025zsi}, complemented by precise electron-scattering data~\cite{Ankowski:2022thw} and direct measurements on ejected neutrons~\cite{MINERvA:2019wqe,MINERvA:2023ikp,MicroBooNE:2024hun,Munteanu:2019llq,Dolan:2021hbw}. Additionally, the proposed nuSCOPE experiment at CERN~\cite{Acerbi:2025wzo} offers event-by-event knowledge of \textit{neutrino} energy and so would allow an almost direct measurement of the neutrino energy bias shown in \autoref{fig:nurecdemo}, covering the full range of energies of interest to DUNE.

%%% FSI for HK
Whilst the case study shown uses the DUNE flux and baseline, the qualitative conclusions are likely to be broadly applicable to any neutrino oscillation experiment relying on the calorimetric neutrino energy reconstruction implied by \autoref{eq:enuhad}. Conversely, the conclusions of this study cannot be directly translated to other methods of neutrino energy reconstruction such as those used by HK, where FSI effects intervene primarily by altering sample composition via pion absorption~\cite{NuSTEC:2017hzk}. %which primarily rely on lepton kinematics. %Even so, FSI effects remain a major uncertainty for such measurements by altering the composition of event samples through pion absorption processes~\cite{NuSTEC:2017hzk}. %The use of water Cherenkov detectors necessitates a less precise neutrino reconstruction method which primarily relies on inferring the neutrino energy from final-state lepton kinematics under the assumption of a quasi-elastic two-body scatter. FSI effects are still relevant for this method, but intervene mostly through the modeling of pion absorption processes which may contaminate quasi-elastic-like samples~\cite{NuSTEC:2017hzk}. 

\textit{Conclusions} ---
In conclusion, this study quantitatively demonstrates that FSI modeling uncertainties have the potential to alter reconstructed neutrino energy spectra in ways which are degenerate with variations of neutrino oscillation parameters. The variations in reconstructed energy spectra due to different FSI models are comparable to, or even exceed, the projected precision of next-generation experiments on $\Delta m_{32}^2$ and $\delta_{\rm CP}$. Whilst this is not directly illustrative of the size of FSI-related systematic uncertainties on oscillation measurements it highlights a critical need for improved FSI modeling, including a comprehensive uncertainty quantification to permit robust near detector constraints, supported by dedicated experimental measurements.% Future work may focus on conducting dedicated measurements of hadron-argon interactions, understanding disagreement among different FSI models, and implementing robust FSI uncertainty parameterisations to support the ambitious physics goals of next-generation oscillation experiments. 

%%%%%%%%%%%%%%%%%%%%%%%%%%%%%%%%%%%%%%%%%%%%%%%%%%
\begin{acknowledgments}

\textit{Acknowledgments} --- YL is supported by the National Science Foundation under
Grant No. 2514289, and would like to thank Edward Blucher for fruitful discussions on this study. All authors would like to thank Steven Dytman and Richard Diurba for discussions related to this work. All authors would also like to acknowledge support from the DUNE collaboration, and the suggestions provided by Julia Tena Vidal, Orlando Peres and Vivek Jain during the review of this work. 
\end{acknowledgments}

%\iffalse

\clearpage
\section{Supplemental Material}

\subsection{Details of FSI models considered}
\label{app:fsi_models}

%his section gives an overview of the FSI models considered within this work. Further details can be found in Ref.~\cite{Dytman:2021ohr}.

This section contains a summary of the FSI models considered in this work.% alongside ...

\paragraph{hA2018} -- This is an empirical model in which hadrons are transported through the remnant nucleus in a single step, allowing them to undergo one simulated interaction. Although some information about this model can be found in the GENIE Physics and User Manual~\cite{genie_manual}, the documentation is incomplete.
The model assigns a probability for different ``fates'' for the hadrons produced at the interaction vertex. For example, hadrons can undergo elastic interactions, be absorbed, knock out other nucleons or undergo other inelastic interactions. The  probabilities for each fate have been tuned to pion- and proton-nucleus scattering data where available (mostly for hadron momenta below 300 MeV/$c$, for example using Ref.~\cite{Ashery:1981tq}), and extrapolated or tuned to other model predictions at higher energies (namely the CEM03 calculations~\cite{Mashnik:2005ay}). Despite being a relatively simplistic model, hA2018 has the advantage of being fully reweightable, which makes it an appealing choice for experimental productions. It is, for example, the baseline model used by the MINERvA~\cite{MINERvA:2025tem}, DUNE~\cite{DUNE:2021mtg}, and SBN experiments~\cite{MicroBooNE:2016pwy,Machado:2019oxb,ICARUS:2004wqc}. 

\paragraph{hN2018} -- This is an full intra-nuclear cascade (INC) model. Its philosophy is very similar to that of the NEUT~\cite{Hayato:2021heg} and NuWro~\cite{Zmuda:2015twa} generators' cascade models. This is a so-called ``space-like'' INC, meaning it propagates one hadron through the nucleus at a time in discrete steps, evaluating its interaction probability at each one, which depends on its position inside the nucleus. At each step hN2018 can simulate elastic and inelastic scattering, charge exchange, pion production and pion absorption. It does not, however, simulate nuclear cluster production. For pions with kinetic energies below 350 MeV, hN2018 uses the Salcedo-Oset model~\cite{Salcedo:1987md} to calculate scattering probabilities; whilst above this, it instead uses free $\pi N$ cross-section measurements. Nucleon reinteractions are based on nucleon-nucleon cross section measurements fit in the partial wave analyses of the GWU group~\cite{said_nn_pwa}. Nucleon interactions take into account medium modifications as prescribed in Ref.~\cite{Pandharipande:1992zz}. 

\paragraph{INCL} -- The INCL++ model~\cite{Mancusi:2014eia,mancusi2014extension, Boudard:2012wc} is an INC model which, unlike the other models, uses a ``time-like'' cascade rather than a space-like cascade to transport ensembles of particles through the remnant nucleus, in steps of space and time, all together. The projectile particle and all the particles in the residual nucleus propagate freely in a square-well potential whose width depends on its location inside the nucleus and the nucleon kinetic energy. The INCL model is based on a classical description of the nucleus, with modifications to account for quantum effects such as Pauli blocking, binding energy, and the non-zero probability of nucleons going past the boundary of the square-well potential. The model contains several scattering channels which are not present in most neutrino generator INC routines, such as nuclear cluster emission. Once the maximum allowed time for the cascade process is reached, the nuclear state is passed to the pre-equilibrium de-excitation routine ABLA~\cite{Kelic:2009yg}, which can produce additional particles or nuclear fragments. A discussion of the performance of the standalone INCL model, coupled with the ABLA routine, compared to the NuWro FSI cascade can be found in Refs.~\cite{ershova2022study,Ershova:2023dbv}. With respect to hA2018 and hN2018, INCL++ is a more sophisticated model due to its more careful description of the nuclear medium, inclusion of quantum effects and simulation of additional fates. It has also shown remarkable agreement with a variety of measurements, including spallation data~\cite{David:2015ura,leray_2011_0qrsx-3rn11}. The current GENIE implementation of INCL is known not to replicate all physics of the original code. Work is ongoing to improve the implementation. 

\paragraph{GENIE with GEANT4 Bertini Cascade (G4BC)} -- The Bertini Intranuclear Cascade model~\cite{Wright:2015xia} is a classical model solving on average the Boltzmann equation for the transport of a particle through a gas of nucleons, which is an adequate approximation if the effective nucleon size is small and there are few collisions. The model is used within the GEANT4 package, and the GENIE generator can interface with it directly through a dedicated package as described in Ref.~\cite{genie_g4bc}. This is a space-like cascade (like hN2018), but contains different choices in terms of the modeling and tuning of physics parameters. For example, unlike the hN2018 model, it includes Pauli blocking. The elementary process cross sections were obtained from CERN-HERA data compilations (Refs.~\cite{Flaminio:1983cz,Flaminio:1983fw,Flaminio:1984gr,HERAGroup:1987dng}) and SAID analyses (Ref.~\cite{said_nn_pwa}). Nucleons produced in these reactions can be grouped into clusters if their momenta are close enough, following relative momentum values from Ref.~\cite{Toneev:1983cb}.

\paragraph{FSI model spread as an uncertainty envelope}

All of the models make different choices in the way they simulate hadron transport, yield different sets of final states and are tuned to different sets of measurements. Although it is the most simplistic, the hA2018 model has been extensively tuned to hadron scattering measurements, and is therefore by definition able to make reasonably accurate predictions where the data sets it was tuned on are applicable. More sophisticated models such as hN2018, INCL and G4BC instead attempt to model the physics of hadron transport in a more sophisticated way, but each relies on different approximations and tunes to different data sets.  

To assess their performance, these models can be benchmarked against hadron scattering data. However, results of such comparisons are inconclusive. For example, the exclusive measurement of pion-argon scattering in Ref.~\cite{DUNE:2025zhx} rejects the hA2018 model and indicates a preference for the G4BC model, and does not rule out the hN2018 and INCL models. Conversely the inclusive measurement in Ref.~\cite{DUNE:2025pda} rejects all models other than the G4BC model. Note that these measurements are able to provide constraints only in the high momentum region (pion momenta above 400 MeV/$c$) where models are most similar. At lower momenta, the model predictions diverge significantly. The divergence of model predictions in regions of momenta lacking data is also clearly demonstrated in Ref.~\cite{PinzonGuerra:2018rju}. 

Beyond differences in their prediction of hadron scattering cross sections, models also differ substantially in the way they treat the relationship between hadron scattering and FSI in neutrino or electron scattering, as discussed at length in Ref.~\cite{Dytman:2021ohr}. Unfortunately, benchmarking hadron transport models directly using neutrino or electron scattering data is challenging due to the lack of precise control over the initial state hadron and the complicated variety of other nuclear effects at play. Electron scattering measurements are able to constrain broad properties of hadron transport models such as transparency (see e.g. Refs.~\cite{CLAS:2025fqh,ershova2022study,Dytman:2021ohr, Niewczas:2019fro}), but this does not constrain the detailed kinematics of hadron transport and such measurements give no conclusive preference for one of the considered models over others. Whilst neutrino scattering measurements can be sensitive to FSI, inference of meaningful constraints is even more challenging due to the unknown incoming energy making an even more acute degeneracy between FSI and other nuclear effects (see e.g. Ref.~\cite{Filali:2024vpy}).

Overall, all the available models have known approximations and comparisons of their predictions to available data does not result in a clear preference for one of them. As a result, the spread of predictions between the four models can be considered a reasonable minimal uncertainty envelope for this work.

\subsection{Impact of FSI modeling on $\sin^2\theta_{23}$}
\begin{figure}[htpb]
\centering
\includegraphics[width=0.40\textwidth]{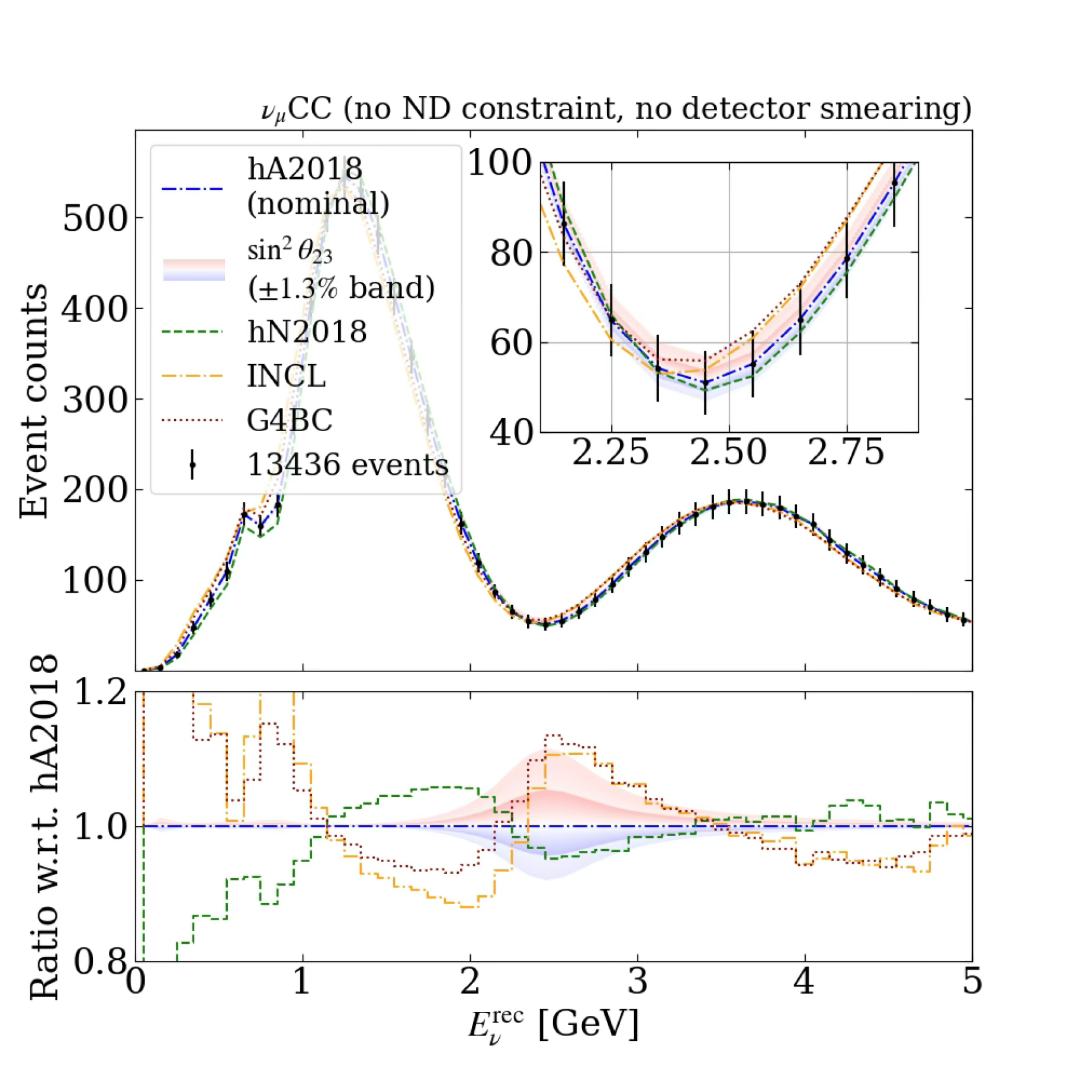}
\includegraphics[width=0.40\textwidth]{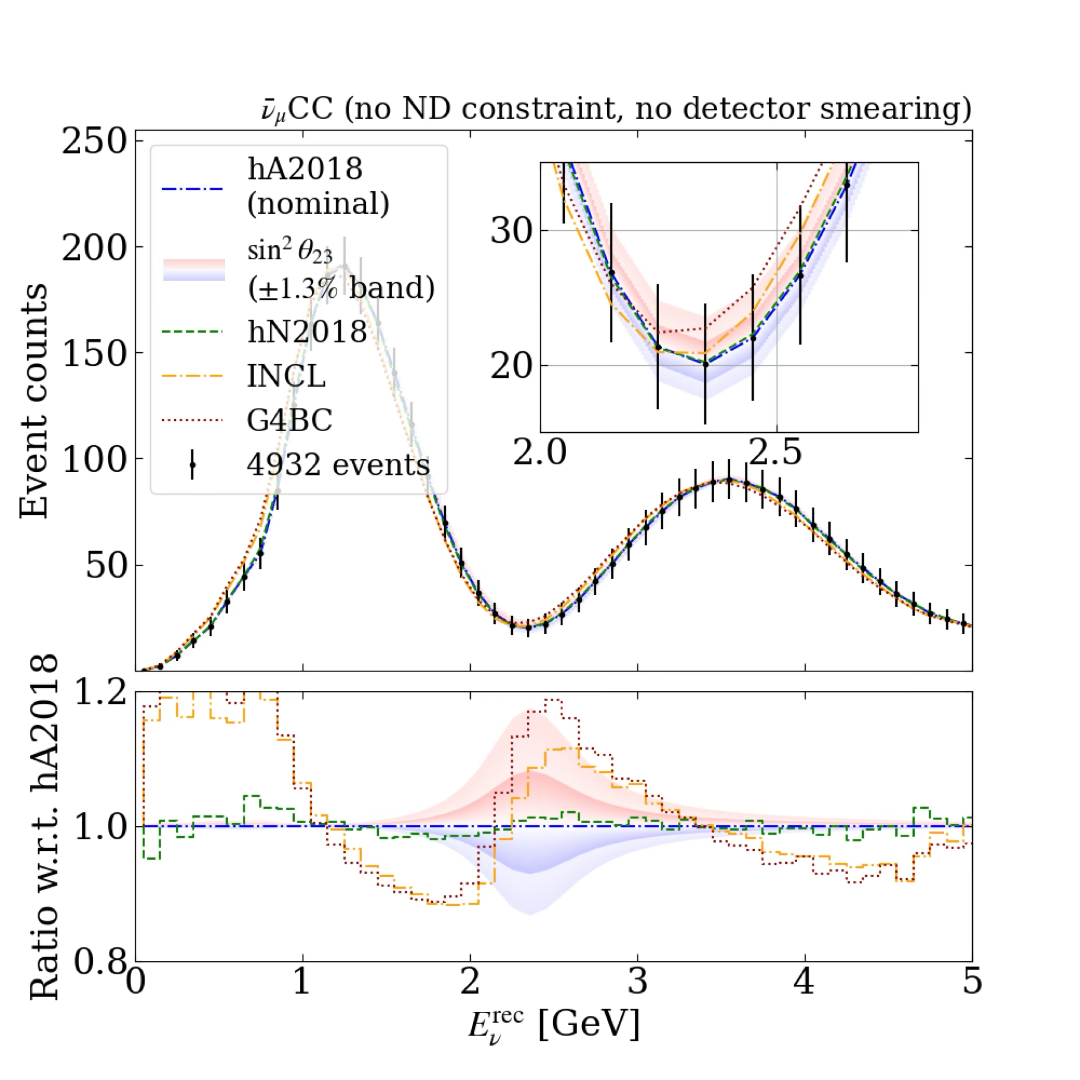}
\caption{Comparison of the simulated $\nu^{\bracketbar}_{\mu}$ reconstructed energy spectra and an oscillated DUNE experiment flux using different FSI models, to $\pm 1.3$\% variations of $\sin^2\theta_{23}$. This is simulated for neutrino (top) and antineutrino (bottom) beam modes. Other figure elements follow the same conventions as described in \autoref{fig:numu_dm32}.} 
\label{fig:numu_st32}
\end{figure}

In addition to the studies presented in the main text, we have tested the effect of FSI modeling on the measurement of the $\sin^2\theta_{23}$ parameter. This is illustrated in \autoref{fig:numu_st32}, where comparisons of FSI model variations with $\pm 1.3$\% variations of $\sin^2\theta_{23}$ (chosen to be indicative of the target precision of next-generation experiments) show that the former can be disambiguated from the latter. This is because changes in $\sin^2\theta_{23}$ mainly affect the normalization of the spectra in the region of the oscillation dip whilst variations of FSI tend to shift the distribution. 

\subsection{Studying the role of detector smearing and thresholds}

\begin{figure}[htpb]
\centering
\includegraphics[width=0.40\textwidth]{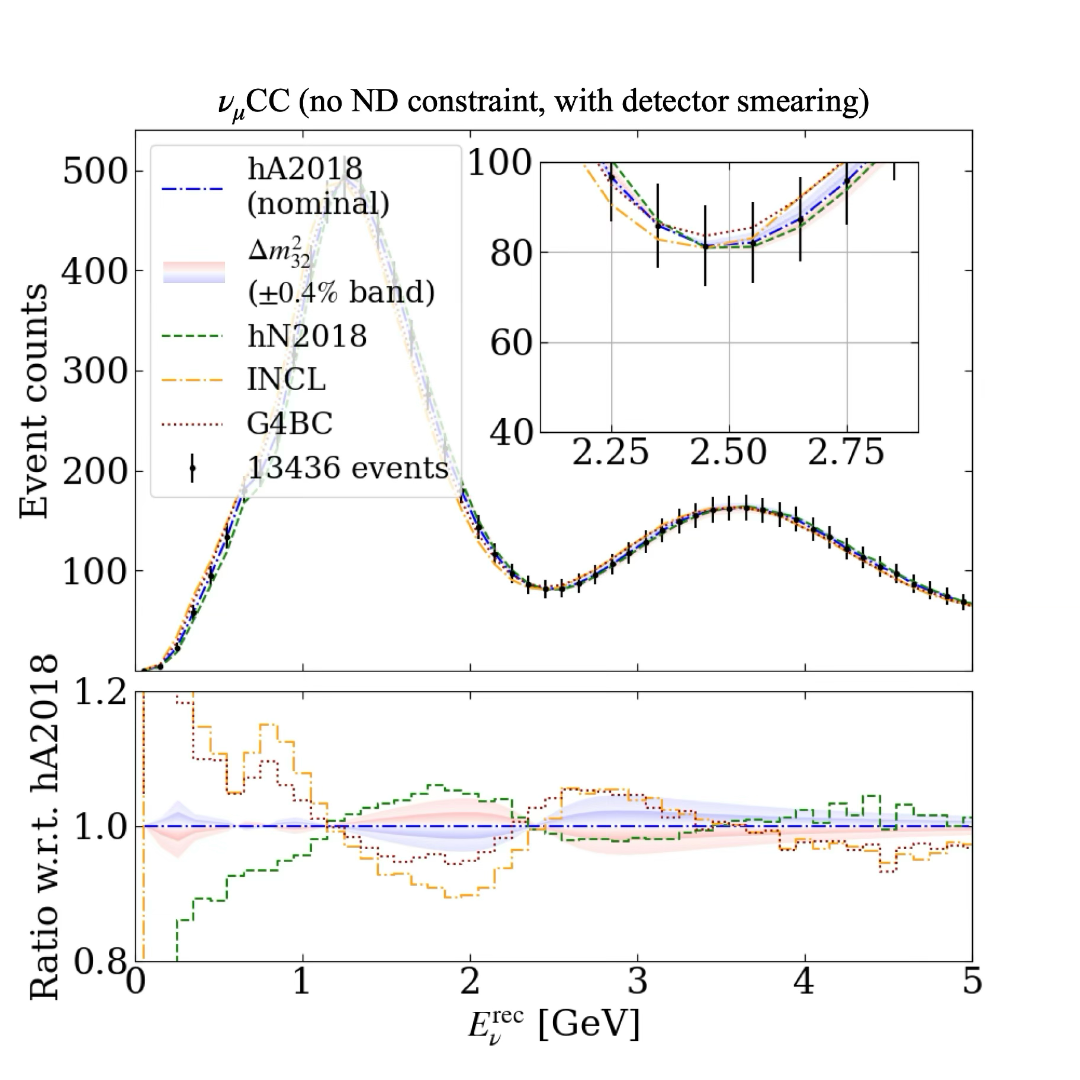}
\includegraphics[width=0.40\textwidth]{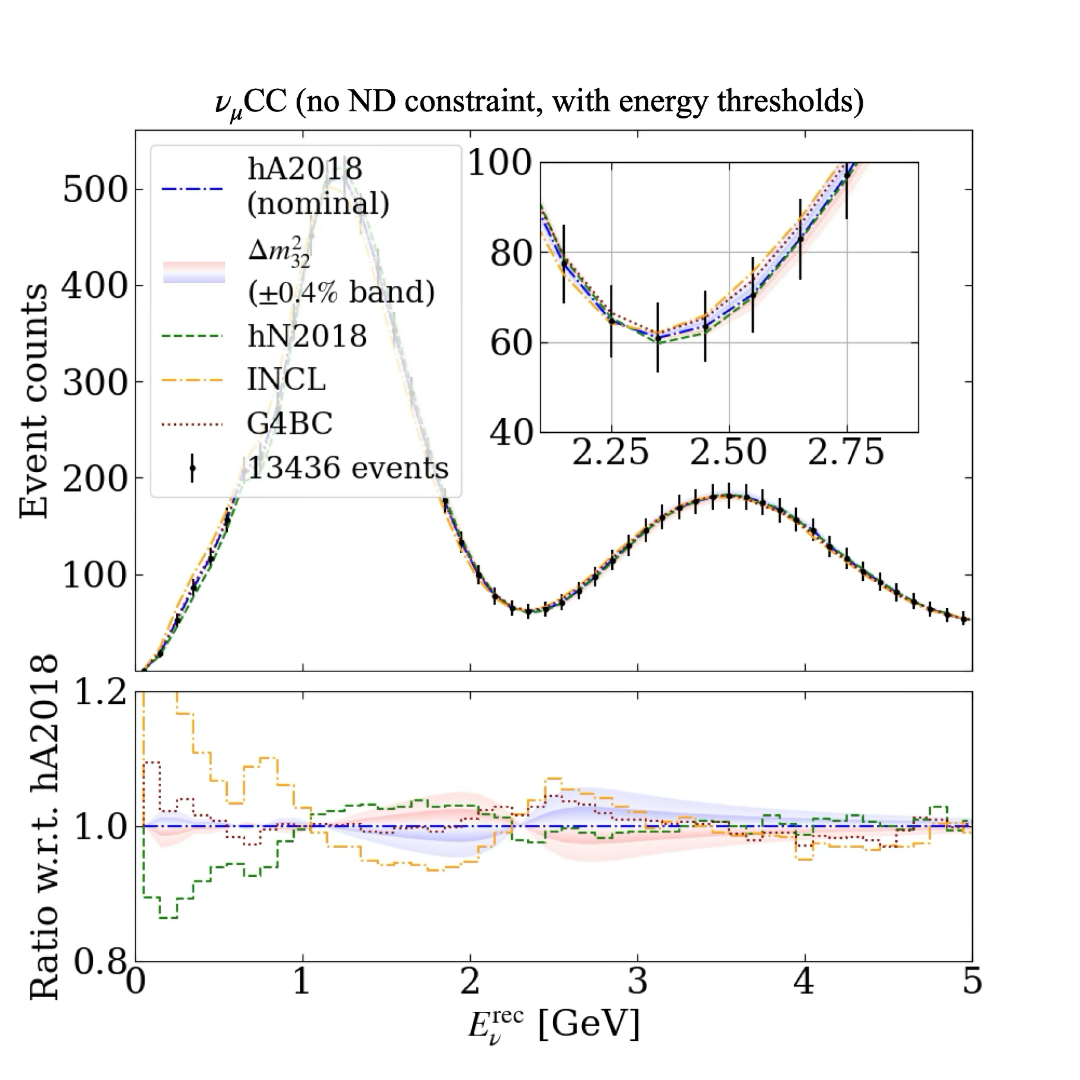}
\caption{Comparison of the simulated $\nu_{\mu}$ reconstructed energy spectra and an oscillated DUNE experiment flux using different FSI models, to $\pm 0.4$\% $\Delta m_{32}^2$. This is simulated including a 34\% Gaussian smearing of the hadronic (top) or including a 100 MeV threshold on outgoing hadron kinetic energy (bottom). Other figure elements follow the same conventions as described in \autoref{fig:numu_dm32}.} 
\label{fig:deteffects}
\end{figure}

We studied the impact of how a simple implementation detector smearing and reconstruction thresholds can alter the studies presented within this work. For the former we considered a 34\% Gaussian smear to the hadronic energy when building $E^{\mathrm{rec}}_{\nu}$ in \autoref{eq:enuhad}, taken from Ref~\cite{DUNE:2020jqi}. For the latter we consider ignoring all hadrons with a kinetic energy of less than 100 MeV when building $E^{\mathrm{rec}}_{\nu}$ (representing a very conservative threshold for a LArTPC detector). Whilst this does change the shape of $E^{\mathrm{rec}}_{\nu}$ and the impact of FSI or oscillation parameter variation, the broad conclusion that variations of the former can be comparable to or larger than variations of the latter remains unchanged. As an example \autoref{fig:deteffects} shows a comparison of variations of FSI models to variations of $\Delta m_{32}^2$ in the presence of the assumed detector smearing and thresholds.

\subsection{The sensitivity of the energy dependence of the reconstructed neutrino energy bias to FSI modeling}

In the main text, \autoref{fig:nurecdemo} shows that FSI changes the magnitude and shape of the bias in our proxy for the neutrino energy reconstruction at DUNE. Changing the FSI model varies the number of (anti)neutrino events with more than 10\% relative bias by about 30\% (20\%). %Whilst not explicitly shown here, the evolution of the bias with neutrino energy is also changed (see supplementary material).

\begin{figure*}[hptb!]
\centering
\includegraphics[width=0.32\textwidth]{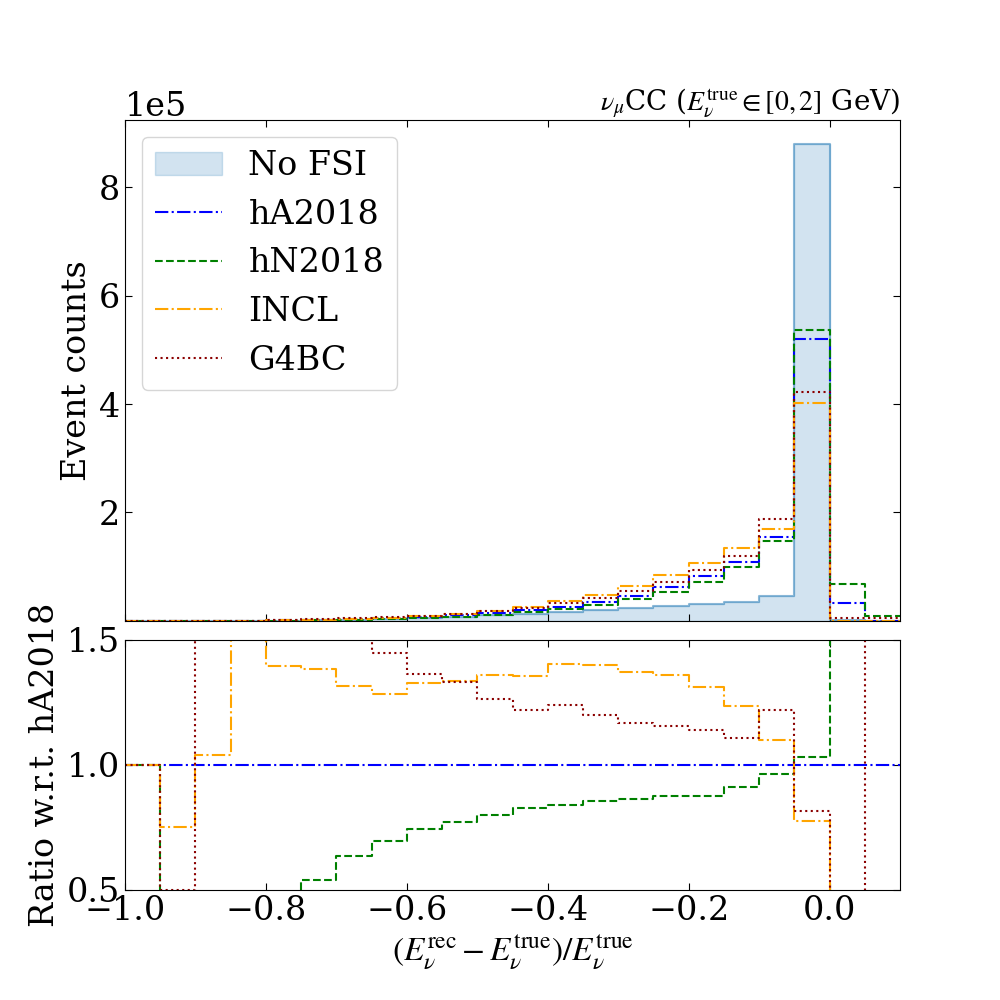}
\includegraphics[width=0.32\textwidth]{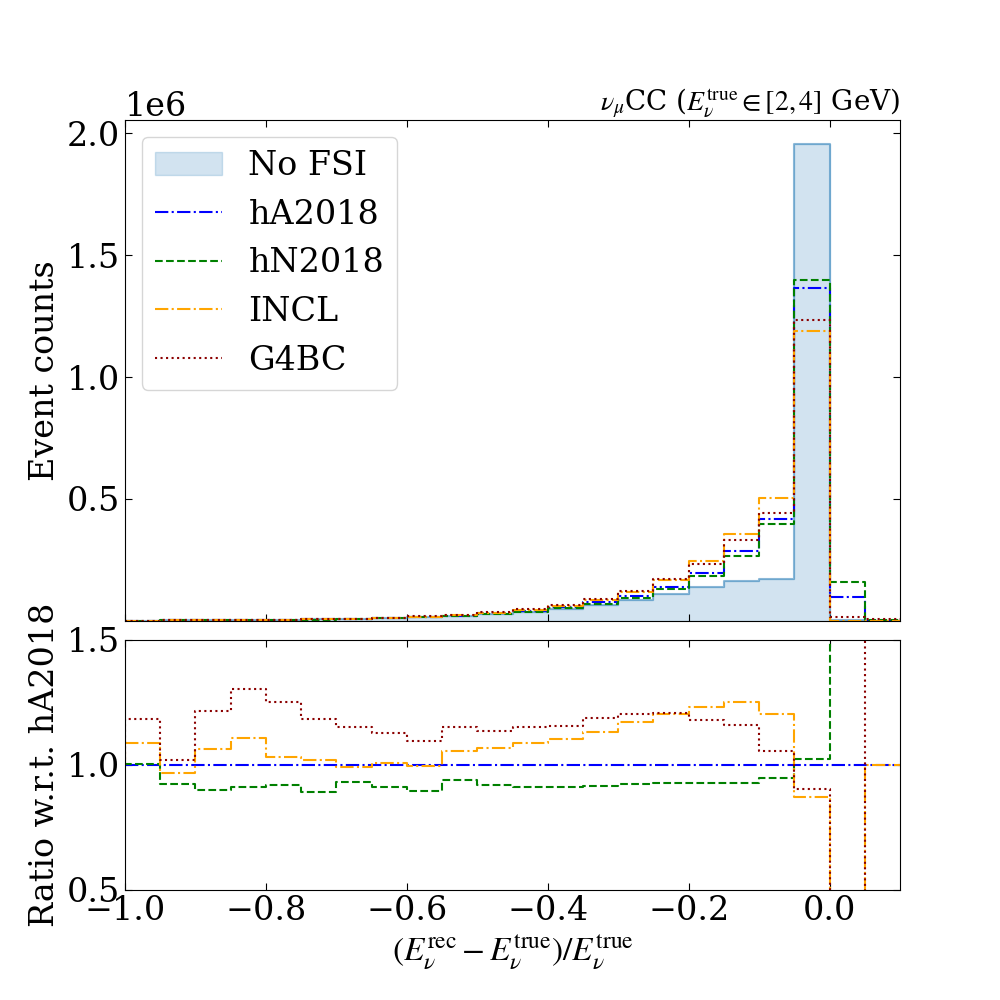}
\includegraphics[width=0.32\textwidth]{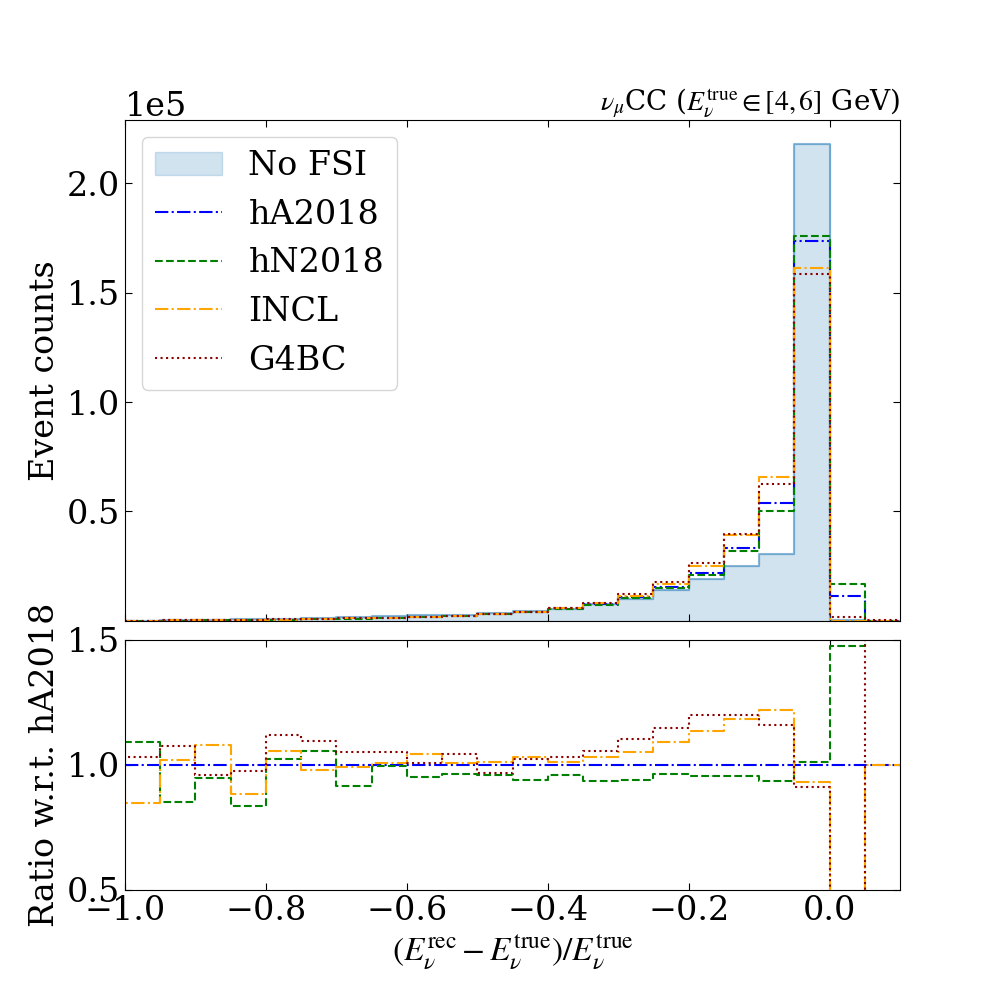}
\includegraphics[width=0.32\textwidth]{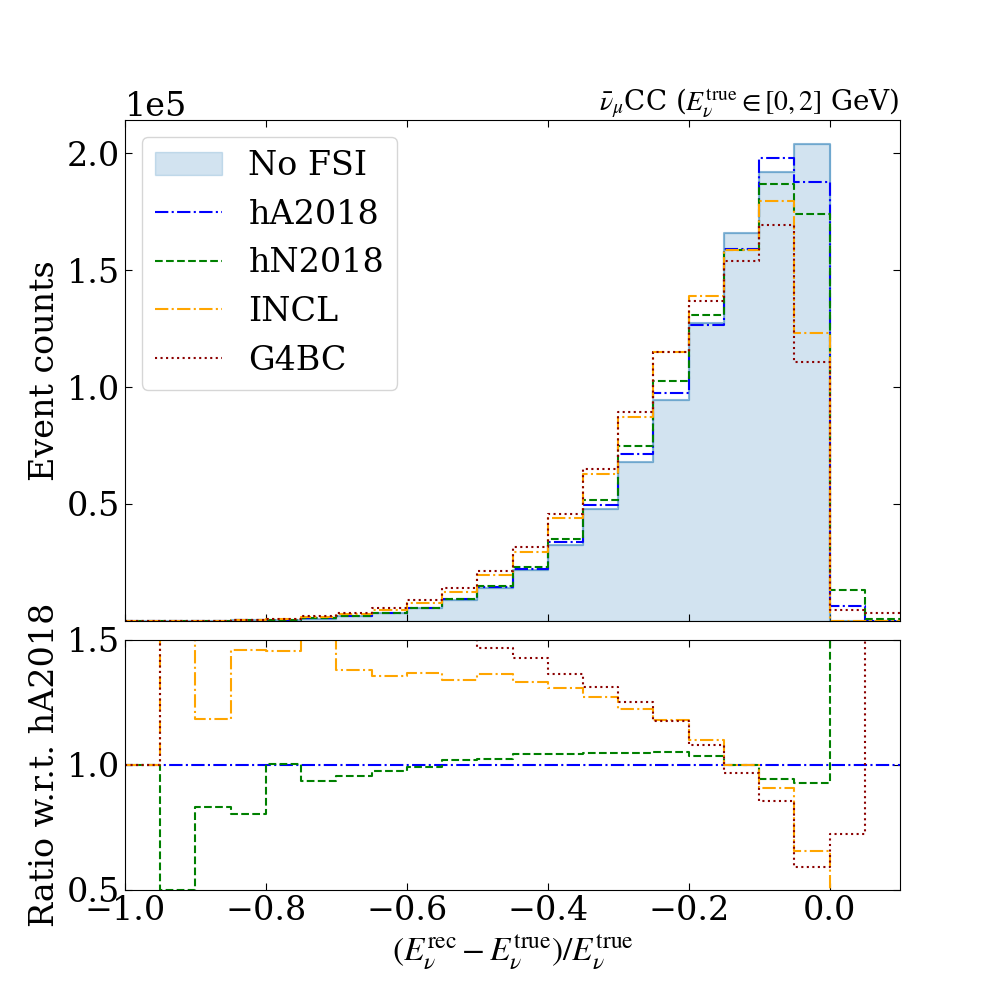}
\includegraphics[width=0.32\textwidth]{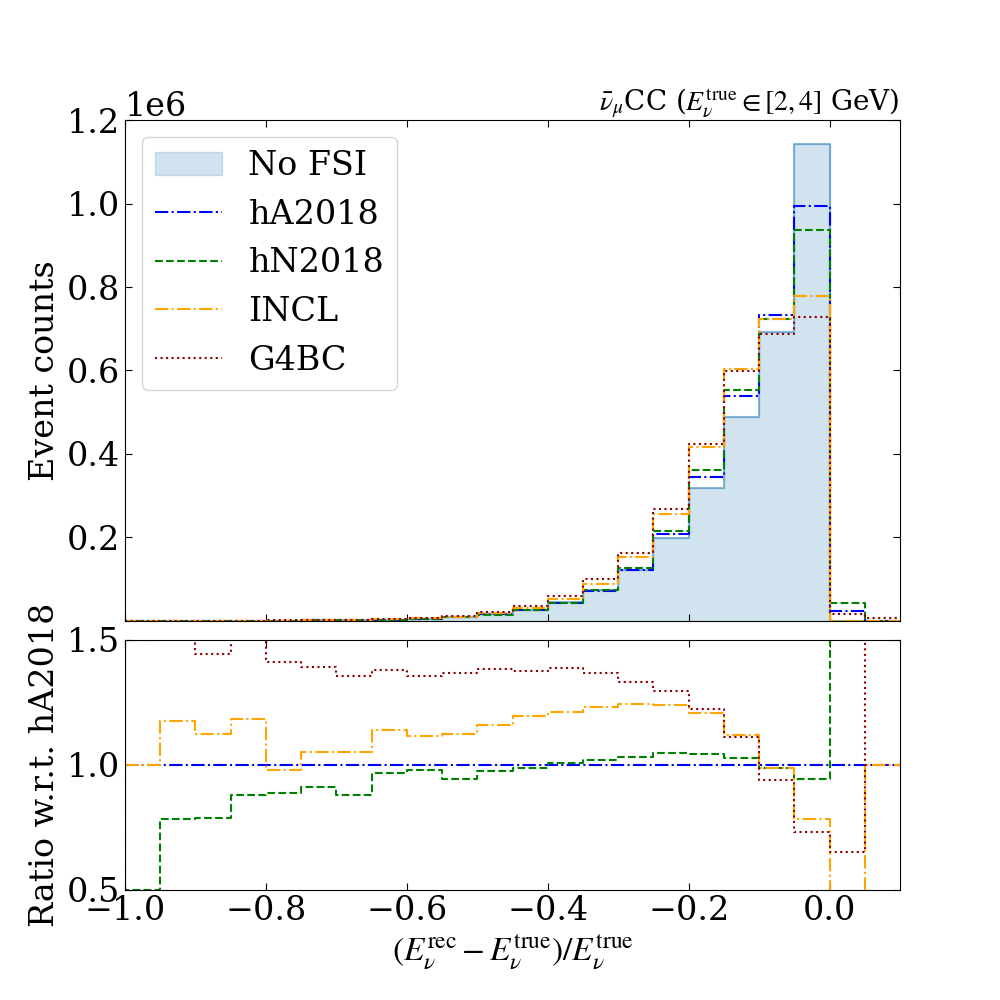}
\includegraphics[width=0.32\textwidth]{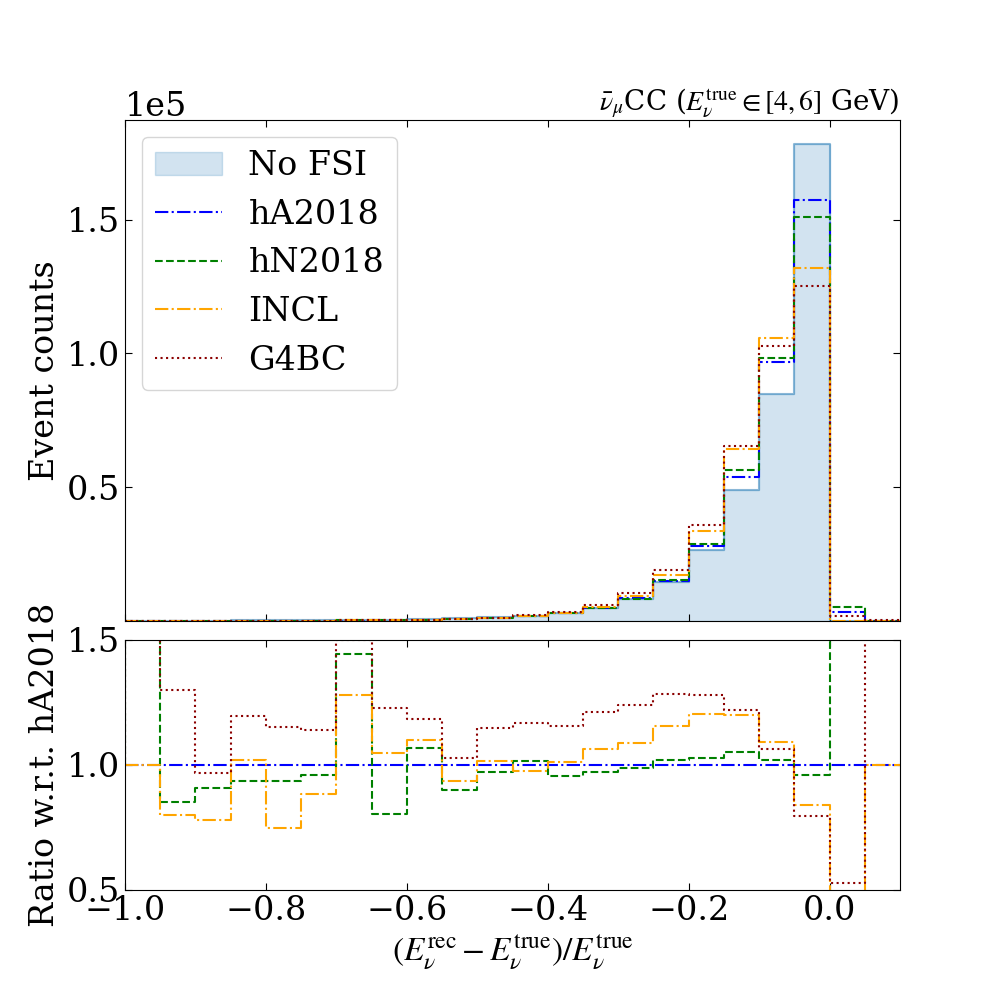}
\caption{The simulated distribution of relative neutrino energy bias for 5 million muon neutrino (top) and antineutrino (bottom) charged-current interactions in three slices of neutrino energy, using the DUNE neutrino and antineutrino beam-mode fluxes respectively before oscillations occur, for a variety of different FSI models. The blue shaded histogram assumes no FSI, while the other histograms correspond to the four FSI models considered in this work. The bottom panel shows the ratios of different FSI model predictions with respect to the hA2018 model. } 
\label{fig:nurecdemoEdep}
\end{figure*}

\autoref{fig:nurecdemoEdep} demonstrates that FSI not only changes the magnitude and shape of the bias in our proxy for the neutrino energy reconstruction at DUNE but also changes the way the bias evolves as a function of neutrino energy. Specifically, \autoref{fig:nurecdemoEdep} shows that at lower neutrino energy the distribution of the neutrino energy bias is more sensitive to differences in FSI modeling than at high neutrino energy.

\subsection{Difficulties with approximating a near-detector constraint}

In this work we consider variations between FSI models as a proxy for their potential effect, neglecting any near-detector constraint. While such constraints are essential in realistic oscillation analyses, simulating a meaningful near-detector constraint in the context of this work is challenging. Any near-detector constraints on FSI  depend on the size and parameterisation of associated uncertainties. At present no experimentally validated uncertainty parameterisation exists that considers the full range of FSI-induced modifications to hadronic energy. %Measurements that are sensitive to FSI effects are not consistently described by any model (see e.g. Refs.~\cite{Filali:2024vpy,Avanzini:2021qlx,CLAS:2025fqh}).

As a result of lacking a well-motivated uncertainty parameterization for FSI, any simulated near detector constraint essentially builds in its own answer. If FSI is parameterized in a way that correlates near detector observables with missing energy, the near detector will provide tight constraints. If FSI is parameterized with freedom to vary missing energy in a way that is uncorrelated with near detector observables then it will not. Neither choice is uniquely justified today. 

The purpose of this work is therefore not to estimate the plausible size of FSI uncertainties on the far-detector spectra given a constraint from a capable near detector, but rather to illustrate the potential size and impact of such uncertainties if not appropriately constrained. This work motivates the need for improved FSI uncertainty parameterisations and dedicated new measurements to enable the development of reliable near-detector constraints in future analyses. 
%\fi

\bibliography{biblo}

\end{document}